\documentstyle[sprocl,epsf]{article}

\def\be{\begin{equation}}
\def\ee{\end{equation}}
\def\bea{\begin{eqnarray}}
\def\eea{\end{eqnarray}}
\begin{document}

\title{MACHOS IN THE SOLAR NEIGHBOURHOOD}

\author{ B. FUCHS, H. JAHREISS}

\address{Astronomisches Rechen--Institut Heidelberg,\\
M\"onchhofstr.~12--14, 69120 Heidelberg, Germany}

\maketitle
\abstracts{We rediscuss the question of what fraction of the population
of MACHOs in the halo of the Milky Way might be in the form of halo dwarfs,
because recent determinations indicate masses of the MACHOs, which are typical
for such stars. For this purpose we have analyzed the Third Catalogue of Nearby
Stars and identified the subdwarfs and a high velocity white dwarf in the solar
neighbourhood. The local mass density of these stars is 1.5$\cdot$10$^{-4}$
$\cal M_\odot$ pc$^{-3}$, which is only 3\% of the current estimate of the
local mass density of the MACHO population. We compare the local density of
subdwarfs with constraints set by HST observations of distant red dwarfs.
Using models of the stellar halo with density laws that fall off like
$r^{-\alpha}$, $\alpha$ = 3.5 to 4, we find that the HST constraints can only
be matched,
if we assume that the stellar halo is flattened with an axial ratio of about
0.6. The non--detection of the analogs of MACHOs in the solar neighbourhood
allows to set an upper limit to the luminosity of MACHOs of $M_B >$ 21
magnitudes.}

\section{Introduction}
The MACHO \cite{alk} and EROS \cite{ans} collaborations have reported new
results of
their campaigns of observing micro-lensing events towards the LMC. These
indicate that
MACHOS have masses typically of half a solar mass and contribute about one half
to the mass budget of the halo of the Milky Way (see the article by
Bennett in this volume). Such masses are typical for red and white dwarfs.
Bahcall and collaborators \cite{bah}$^{\!,\,}$\cite{fly} have made by very
deep star counts based on HST data {\em in situ} measurements of the space
density of such stars in regions, where the micro-lenses are expected to be
physically located. Their conclusion is that halo dwarfs
contribute only a few percent to the mass of the halo and are thus unlikely
MACHO candidates. Graff \& Freese \cite{gra} (see also Freese's article in this
volume) infer from the same data that the local
mass density of halo dwarfs is even less than one percent of the combined local
densities of the dark and stellar halos.

On the other hand, halo dwarfs are directly observed in the solar
neighbourhood.
Liebert \cite{lie} discusses the results of the USNO parallax programme for red
dwarfs and
finds that the number of halo dwarfs in that sample is consistent with the HST
data. We have followed a complementary approach. Recently the Third Catalog of
Nearby Stars (CNS3) \cite{jah} has been completed at the Astronomisches
Rechen-Institut.
\begin{figure}[htb]
\epsffile{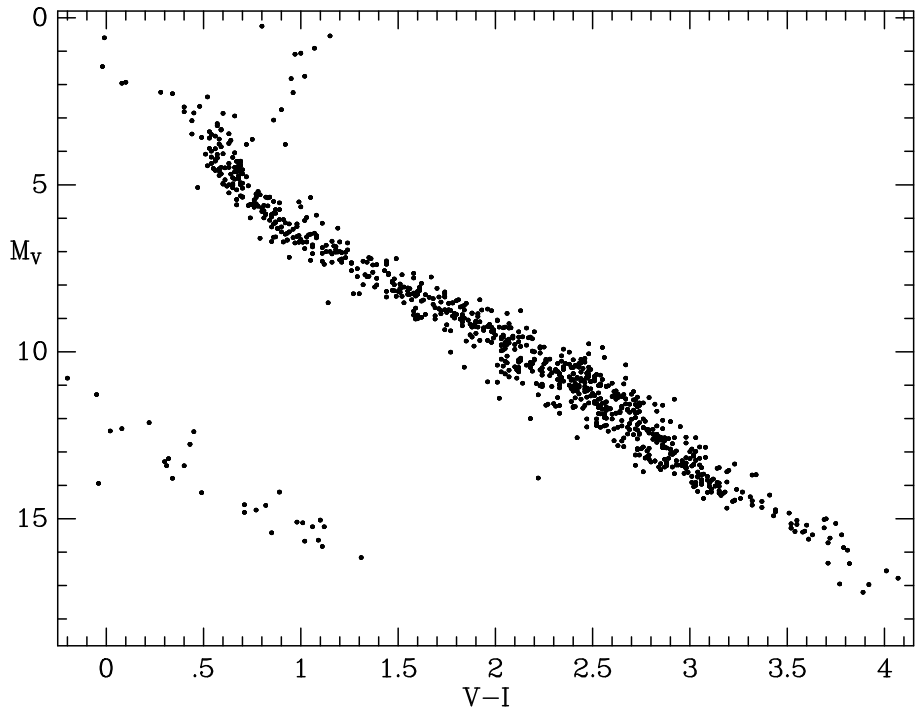}
\caption{Colour--Magnitude diagram of nearby stars with $\sigma_{M_V} \leq$
0.3 mag.}
\end{figure}
This provides the now most complete inventory of the solar neighbourhood up to
a distance of 25 pc from the Sun and allows a new determination of
the local density of halo stars.
\begin{table}[htb]
\caption{Subdwarfs in the CNS3.}
\vspace{0.4cm}
\begin{center}
\begin{tabular*}{12.0cm}{|l@{\extracolsep\fill}@{\hspace{0.3cm}}rrrrrrrc|}
\hline
\rule[0mm]{0mm}{4mm}
  & \multicolumn{1}{c}{d} & \multicolumn{1}{c}{$M_V$}   & \multicolumn{1}{c}{
       $V-I_c$} & U  & \multicolumn{1}{c}{V}  & W  & $[$Fe/H$]$ &
\multicolumn{1}{c|}{$\cal M$}\\[0.5ex]
        & [pc]  & [mag] & \multicolumn{1}{c}{[mag]} &  & [km/s] &  & & [$\cal
M_\odot$]\\[0.5ex]
\hline
Gl\,191  & 3.9  & 10.90   & 1.98  & 19  & --288  & --54 &$<-$1 & 0.2\\
[0.5ex]
Gl\,299  & 6.8  & 13.65   & 2.83  & 107 & --126  & --45 &$<-$1 &
0.1\\[0.5ex]
Gl\,53A & 7.5 & 5.80  & 0.79 & --42 & --157 & --34  & --0.6 & 0.73 \\[0.5ex]
Gl\,53B & 7.5 & 11.00 & & & & &  &  0.1 \\[0.5ex]
Gl\,451A & 8.9 & 6.70  & 0.85 & 273 & --154 & --16  & $<-$1 & 0.6 \\[0.5ex]
Gl\,699.1 & 15.6  & 13.34   & 0.50  & --149  & --294  & --42 && 0.6
\\[0.5ex]
GJ\,1062 & 16.0  & 12.00   & 2.76  & 93  & --212  & 17 & $<-$1 &
0.15\\[0.5ex]
Gl\,781 & 16.5 & 10.91  & 2.56 & 104 & --47  & 34 & $<-$1 & 0.2 \\[0.5ex]
Gl\,158 & 18.4 & 7.17  & 0.93 & --41 & --188 & 21 & $<-$1 & 0.6 \\[0.5ex]
LHS\,3409 & 20.3 & 13.59  & 2.76 & --31 & --101 & 44 & low &
0.1\\[0.5ex]
GJ\,1064A & 23.6 & 6.31  & 0.86 & --94 & --109 & --74 & --1 & 0.67\\[0.5ex]
GJ\,1064B & 23.6 & 6.91  & 1.02 & --94 & --111 & --77 & $<-$1 & 0.63\\[0.5ex]
WO\,9722 & 23.9 & 11.39  & 2.05 & 264 & --215 & --92 & low &
0.17\\[0.5ex]
LHS\,375 & 23.9 & 13.78  & 2.27 &\multicolumn{3}{c}{$v_t$ = 158}&  $<$--1 & 0.1
\\[0.5ex]
\hline
\end{tabular*}
\end{center}
\end{table}

\section{Subdwarfs in the CNS3}
We have searched the CNS3 for halo dwarfs. These reveal themselves by their
position on the subdwarf sequence in the colour-magnitude diagram (CMD), their
low metallicity, and their high space velocities. Unfortunately, there is in
this metallicity range a broad overlap of the comparatively few halo stars and
the much more abundant thick disk stars. We have therefore adopted a
very conservative search  strategy in order to isolate a true halo population,
even if this will lead to an undersampling of the halo population. We have
examined first the stars on the subdwarf sequence, which can be clearly seen in
the CMD shown in Fig.~1. The reliability of the photometry and the parallaxes
of each star has been checked carefully, which led to the omission of quite a
number of stars with inaccurate parallaxes.
In this way we identified Gl\,191,
Gl\,781, WO\,9722 = LHS\,64, GJ\,1062, LHS\,375, and LHS\,3409 as likely halo
stars. A
number of stars in the CNS3 are classified as subdwarfs by their spectral type.
This enables us to search for subdwarfs in that part of the CMD, $V-I <$ 1.5,
where the subdwarf sequence comes very close to the main sequence. We include
GJ\,1064A+B and Gl\,53A+B, which have reliable parallaxes, into our list on
the
basis of this criterion. All stars have decidedly large space velocities (cf.
Table 1 and Fig.~2). In addition to these stars we found 4 stars, Gl\,158,
Gl\,299,
Gl\,451A, and the white dwarf Gl\,699.1 (spectral type DA7), which do not lie
very
prominently on the subdwarf sequence, but have space velocities,
which clearly identify them as halo stars.

All the stars in our list have very low metallicities. Leggett \cite{leg} has
determined
the metallicities of a large number of nearby M dwarfs from their position in
the (J--H)--(H--K) two infrared-colour diagram. According to that determination
Gl\,191, Gl\,781, GJ\,1062, and Gl\,299 have metallicities $[$Fe/H$] < -$1.
LHS\,375 has been classified as a cold subdwarf by Ruiz and Anguita \cite{rui}.
The metallicities of Gl\,53A, Gl\,158, Gl\,451,
and GJ1064A, B have been taken from the compilation by Taylor \cite{tay}.
Reid et al.~\cite{rei} have classified GJ\,1062, Gl\,781
\footnote{Gl\,781 is a variable star and shows H$_\alpha$ emission.
However, it is a spectroscopic binary, which, in our view, accounts for this
phenomenon.}, WO\,9722, and LHS\,3409 as likely subdwarfs on the basis of
their spectra. Jones et al.~\cite{jon} confirm the low metallicity
of Gl\,299. There are many apparently low metallicity stars in
the catalogue, which must be thick disk stars. For instance Leggett \cite{leg}
has assigned to 20 stars in the CNS3 a metallicity of $[$Fe/H$]<-$1, of which
only 4 have been finally included in our sample.

In Table 1 we give the list of stars selected
out of the CNS3 as described above, of which we are now reasonably certain
that they are genuine halo stars.
The velocity components of the subdwarfs given in columns 6 to 8 in Table 1 are
referred to the Sun. In order to reduce them to the LSR the solar motion (U, V,
W)$_\odot$ = (+9, +12, +7) km/s has to be added. Note that U points towards
the galactic center. The velocity distribution of our sample is shown  as a
Toomre diagram in Fig.~2. It is consistent with a distribution centered on
V = --220 km/s. No clustering near V = --40 km/s, which would indicate a
contamination
by thick disk stars, is detected. The dispersion of $\sqrt{U^2+W^2}$ is 150
$\pm$ 30 km/s, which is within statistical errors consistent with other
determinations \cite{lay} of the kinematical parameters of halo stars.
\begin{figure}[htb]
\begin{center}
\leavevmode
\epsffile{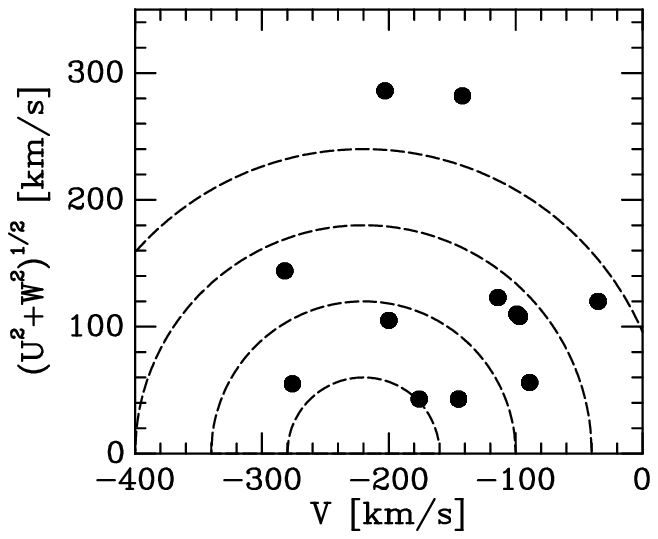}
\caption{Toomre diagram of the nearby halo dwarfs. The dashed lines indicate
lines of constant kinetic energy of the halo stars.}
\end{center}
\end{figure}
The masses given in the last column of Table 1 have been assigned using the
mass-to-luminosity relation calculated theoretically for a metallicity of
$[$Fe/H$]$ = --1.35 by Alexander et al \cite{alx}.

\section{Results and Discussion}
\subsection{Local density}
The local mass density of halo stars can be estimated in various ways. In
Fig.~3
we show density estimates, which have been calculated from the cumulative
distribution of the stars in our sample, i.e.~by adding up the masses of
stars within given distances and dividing by the corresponding spherical
volumes. The stars within 10 pc give rather high density estimates of the order
of 10$^{-3}$  $\cal M_\odot$ pc$^{-3}$. Even though Wielen \& Jahrei{\ss}
\cite{wie} found
similar values, when evaluating the second edition of the Gliese catalogue,
they seem unrealistic to us. We find a plateau in the density run,
when we consider the stars within the 20 pc sphere (cf.~Fig.~3). Beyond that
the density drops off,
probably because the catalogue becomes incomplete. The most likely value of
the local mass density lies according to this determination in the range 1.5 to
1$^.$10$ ^{-4}$ $\cal M_{\odot}$ pc$^{-3}$. The local density of the dark
halo \cite{bss} is about 9$^.$10$^{-3}$ $\cal M_{\odot}$ pc$^{-3}$ (see also
Gates' article in this volume), so that the local density of halo stars is
about 1.7\% of the halo density or 3\% of the local MACHO density as determined
by the MACHO collaboration.
\begin{figure}[htb]
\begin{center}
\leavevmode
\epsffile{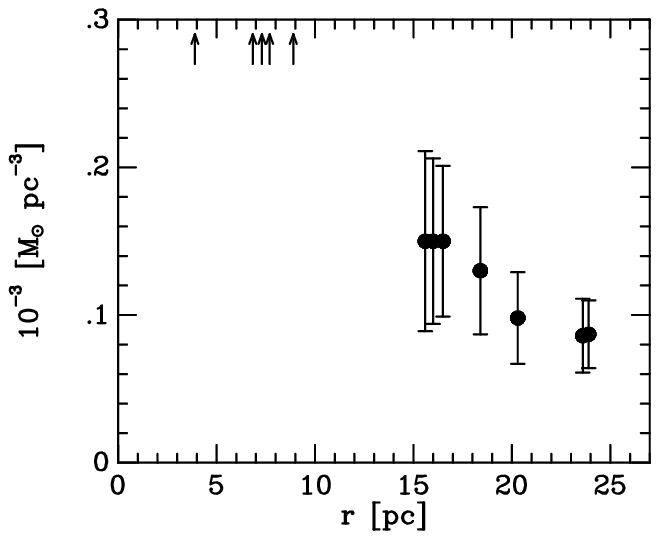}
\caption{Mass density of nearby halo dwarfs as function of the distance from
the Sun. Within 10 pc only the positions of the stars are indicated. The
errorbars indicate the statistical uncerntainties.}
\end{center}
\end{figure}
We have broken down our sample of subdwarfs with respect to absolute
magnitudes. The resulting `luminosity function' has been compared with the
subdwarf luminosity function obtained by Dahn et al.~\cite{dan} for stars with
$M_V >$ 10 mag. Both luminosity functions agree well within statistical errors.

\subsection{Matching the constraints by HST data.}
It is interesting to confront the density of halo stars derived here with
density estimates based on the HST data. Graff \& Freese \cite{gra} find in a
detailed analysis of earlier HST data \cite{bah} a value about five times less
than our estimate, which is partly due to their restriction to red dwarfs
redder than $V-I \approx$ 2 mag.
Whereas these authors tried to infer the local density of halo
stars from the star counts, we reverse the approach and make predictions of
the expected number of stars in the star count field. Very recently Flynn et
al.~\cite{fly} have reported an analysis of the Hubble Deep Field (HDF). The
HDF has a size of 4.4 square arcminutes and is located at $l$ = 126$^\circ$,
$b$ = 55$^\circ$.
The limiting magnitudes are 18 to 26.3 mag in the $I$--band.

In the following we assume that the population of halo stars follows a density
law of the form
\begin{equation}
{\nu_h} = {\nu}_{h_\odot} \left( \frac{r_c^2+r_\odot^2}
{r_c^2+r^2} \right)^\alpha
\end{equation}
with an arbitrarily chosen core radius of $r_c=$ 1 kpc. Such density laws are
typical for other tracers of the halo population, such as RR~Lyr stars
\cite{nn} or
horizontal branch stars \cite{pre}, in the Milky Way or external galaxies
\cite{vdb}. The exponent
$\alpha$  lies typically in the range 3.5 to 4. Note that the dark halo is
usually described by a density law of the form as in equation (1) with $\alpha$
= 2. If the subdwarfs are observed locally at a density $\nu_{h_\odot}$, one
predicts by integrating along the line of sight
\begin{equation}
N = \Omega {\nu}_{h_\odot} \int_{d_{min}}^{d_{max}}s^2 \left(
\frac{r_c^2+r_\odot^2}
{r_c^2+r_\odot^2+s^2-2sr_\odot\cos{l}\cos{b}} \right)^\alpha ds
\end{equation}
stars in the star count field. $\Omega$ is the angular area of the field and
the minimum and maximum distances are determined by the limiting magnitudes. We
concentrate first on the colour range $V-I$ = 1.8 to 3.5 mag. We have 4
stars in that range within 16.5 pc, where we believe our sample to be
reasonably
complete. If we project these into the cone towards the HDF, we obtain the star
numbers summarized in Table 2. The limiting distances have been calculated for
each star individually,
\begin{equation}
d_{\min,\max} = 10^{0.2(I_{\min,\max}+(V-I)-M_V+5)} ,
\end{equation}
where, in order to avoid confusion with disk stars in the HDF, we adopt a lower
limit of $I_{min}$ = 24.6 mag. As can be seen from Table 2 we predict
7 to 11 stars in the HDF, whereas Flynn et al.~\cite{fly} have actually
dectected no star.
\begin{table}[htb]
\caption{Predicted Number of Stars in the HDF and the GS.}
\vspace{0.4cm}
\begin{center}
\begin{tabular*}{8cm}{|c@{\extracolsep\fill}cc|cc|}
\hline
\rule[0mm]{0mm}{4mm}
  & \multicolumn{2}{c|}{$V-I >$ 1.8} &
\multicolumn{2}{c|}{$V-I <$ 1.8} \\[0.5ex]
 $\alpha$ & c/a &n$_{\rm HDF}$ & n$_{\rm HDF}$ & n$_{\rm GS}$ \\[0.5ex]\hline
 3.5 & 1 & 11  & 17 & 265 \\[0.5ex]
 4   & 1 & 7 & 7 & 144 \\[0.5ex]
 3.5 & 0.6 & 4 & 5 & 69 \\[0.5ex]
 4   & 0.6 & 2 & 2 & 32 \\[0.5ex]\hline
\end{tabular*}
\end{center}
\end{table}
Despite the low number of stars involved, this discrepancy
seems to be statistically significant. Several explanations might account for
this
discrepancy. First, the stellar halo might be much more irregular and lumpy
than previously thought. Second, the stellar halo is almost certainly flattened
\cite{nn}$^{\!,\,}$\cite{pre}$^{\!,\,}$\cite{vdb}
with an axial ratio around c/a $\approx$ 0.6.
If we take such a flattening into account in the halo model,
\begin{equation}
{\nu_h} = {\nu}_{h\odot} \left( \frac{r_c^2+R_\odot^2}
{r_c^2+R^2+z^2/(c/a)^2} \right)^\alpha\,\,,
\end{equation}
where $R, z$ denote cylindric coordinates, we predict star numbers in the HDF,
which are statistically consistent with no star seen by Flynn et
al \cite{fly}.

The star counts in the HDF in the colour range $V-I <$ 1.8 mag can be
interpreted in a similar way. Disk stars in this colour range are so bright
that they would have to lie several kpc above the midplane to appear fainter
than $I$ = 22 mag. Thus in order to avoid confusion with disk stars in the HDF,
we consider only stars fainter than $I$ = 22 mag.
In our sample there are 4 stars within 20 pc in this colour
range. Their $V-I$ colours cluster around 0.9 and the white dwarf has $V-I$ =
0.5. Thus we define a colour range 0.5 $< V-I <$ 1.8, in which we compare the
predictions with actually observed numbers of stars.  If the stars of our
sample are
projected into the cone towards the HDF using spherical halo models, we predict
7 to 17 stars depending on the model, whereas Flynn at al.~\cite{fly} have
observed 4 stars. If we assume again a flattening of c/a = 0.6,
the number of expected stars is consistent with the observed number.

Gould et al.~\cite{gou} are presently investigating another field observed by
HST.
This is the so called Groth strip (GS), which has an angular size of 114 square
arcminutes and is located at $l$ = 96$^\circ$ and $b$ = 60$^\circ$. Its
limiting magnitude is $I$ = 23.9 mag. Gould et al.~\cite{gou} have very kindly
made their
preliminary data available to us, so that we were able to make star counts in
this field. Due to the limiting magnitude of $I$ = 23.9 mag even the faintest
stars in the GS with colours redder than $V-I$ = 1.8 could be disk stars. Thus
we focus on the colour range $V-I <$ 1.8 and adopt the same halo zone as in
the HDF. The GS has 66 stars in this zone. If we project the local subdwarfs
into the cone towards the GS, we obtain the numbers of predicted stars
summarized in the last column of Table 2. Again a flattened halo model is a
better fit to the data.

\subsection{Constraints on the absolute magnitude of MACHOs.}
Like previous authors we have {\em not} detected the analogs to MACHOs in
the solar neighbourhood. This raises the question why. All subdwarfs in our
sample appear in the LHS high-proper motion star catalogue. Being halo
objects, MACHOs in the solar neighbourhood must have similar proper motions.
The fact that they have not been found in high-proper motion
surveys allows to set an upper limit to their luminosity. The LHS catalogue
\cite{luy},
for instance, is claimed to be complete for stars down to apparent magnitude
$B$ = 21 mag and with proper motions 2.\hspace{-0.2em}$^{''}$5/a $>\mu>$
0.\hspace{-0.2em}$^{''}$5/a. For stars brighter than $B$ = 10 mag it is claimed
that
all stars with proper motions $\mu>$ 0.\hspace{-0.2em}$^{''}$3/a are contained
in
the catalogue. We assume that the MACHOs are homogeneously distributed around
the Sun and have a gaussian velocity distribution,
\begin{equation}
dn = \frac{\nu_{M_\odot}}{(2\pi)^{3/2}\sigma_M^3} exp - \frac{1}{2 \sigma_M^2}
(U^2+(V-\overline{V})^2+W^2) d^3vd^3r\,\, ,
\end{equation}
centered on $\overline{V}$ = --220 km/s and with a velocity dispersion typical
for halo objects, $\sigma_M = \overline{V}/\sqrt{2}$. Integrating now over all
radial
velocities and the proper motions according to the specifications of the LHS we
can determine the radius $r_2$ of the sphere out of which one would expect say
2 MACHOs in the LHS,
\begin{eqnarray}
2 & = &4.74^2 \frac{\nu_{M_\odot}}{\sigma_M} \int_0^{r_2}dr r^4
\int_{\mu_l}^{\mu_
h}d\mu \mu \int_0^{2\pi}dl \int_{-\pi}^{+\pi}db \cos b  \nonumber \\
&  & \cdot I_0\left(\frac{\overline{V}}{\sigma_M^2} 4.74 \mu r
\sqrt{1-\sin^2l\cos^2b}\,\right) \nonumber \\
&  & \cdot exp
-\frac{1}{2\sigma_M^2}\left(\overline{V}^2(1-\sin^2l\cos^2b)+(4.74
\mu r)^2\right)\,\, ,
\end{eqnarray}
where $I_0$ denotes the Bessel function. According to Poisson statistics two
expected MACHOs would be still consistent with no MACHO actually seen in the
LHS survey.
In Table 3, we give results of numerical integrations of equation (6) assuming
a local density of MACHOs of $\nu_{M_{\odot}}$ = 0.5 $\cdot$ 0.01 ${\cal
M}_\odot$
pc$^{-3}$ / 0.5 ${\cal M}_\odot$ as determined by the MACHO collaboration. A
lower limit of the
absolute magnitude of the MACHOs is then given by $M_B = 21 -5\log{r_2} +5$,
because
MACHOs more distant than $r_2$ will not show up in the LHS, since they are
fainter than
the apparent magnitude limit of the LHS, $B$ = 21 mag. From Table 3 we conclude
that MACHOs are fainter than $M_B$ = 21.2 mag. The southern hemisphere
($\delta < -30^\circ$) and the
galactic belt are not sampled by the LHS survey. Thus one could argue that only
the region  $b \ge 30^\circ$ is properly sampled. This would reduce the numbers
in the second column of Table 3  by a factor of 0.25 and shift the
lower limit of the absolute
magnitude of the MACHOs to $M_B$ = 20.6 mag. If the distribution of MACHOs is
irregular and lumpy, the micro-lensing results could mimic a too high mean
density of MACHOs. If we assume a local MACHO density of only 10\% of the value
determined by the MACHO collaboration, the lower limit of the
absolute magnitude of MACHOs would be $M_B$ = 20.1 mag.
\begin{table}[htb]
\caption{Estimated absolute magnitudes of MACHOs.}
\vspace{0.4cm}
\begin{center}
\begin{tabular*}{7cm}{|c@{\extracolsep\fill}cc|}
\hline
\rule[0mm]{0mm}{4mm}
$r_N$             &      N     & $M_B$\\[0.5ex]
[pc]              &            & [mag]\\[0.5ex] \hline
6                 &     0.31   & 22.1  \\[0.5ex]
9                 &     2.3    & 21.2\\[0.5ex]
12                &    9.4    & 20.6\\[0.5ex]
15                &    27.3    & 20.1\\[0.5ex]
18                &    64.0    & 19.7\\[0.5ex] \hline
\end{tabular*}
\end{center}
\end{table}

If the MACHOs were identified
with faint red dwarfs, their masses would be less than 0.1 ${\cal M}_\odot$ in
contradiction to the estimate by the MACHO collaboration.
As an alternative very faint white dwarfs have been discussed in the
literature as MACHO candidates \cite{sil}. If we use
the theoretically calculated cooling sequence of a DA white dwarf by Wood
\cite{mot} and extrapolate $B-V$ to 2 mag,
we estimate cooling times of 11 to 13 Gyrs for white dwarfs as faint as $M_B$ =
20.1 to 21.2 mag. This would make them as old as globular clusters and
raises the question, whether this allows for reasonable lifetimes of the
progenitors of the white dwarfs.

\section*{Acknowledgements}
We are indebted to G.~Gilmore for encouraging us to this research and C.~Flynn,
K.~Freese, N.W.~Evans, and R.~Wielen for many valuable hints and comments. We
are grateful to J.~Bahcall, C.~Flynn, and A.~Gould for making available to us
the their data on the Groth strip prior to publication.

\end{document}